\documentclass{emulateapj}

\begin{document}

%% one line. However, you may use \\ to force a line break if you
%% desire.

\title{Fishing in Tidal Streams: New Radial Velocity and Proper Motion Constraints on the Orbit of the Anticenter Stream}

\author{C. J. Grillmair \footnote{Visiting Astronomer, Kitt
Peak National Observatory, National Optical Astronomy Observatory,
which is operated by the Association of Universities for Research in
Astronomy (AURA) under cooperative agreement with the National
Science Foundation.}  \footnote{The WIYN Observatory is a joint
facility of the University of Wisconsin-Madison, Indiana University,
Yale University, and the National Optical Astronomy Observatory.}} \affil{Spitzer Science Center,
 1200 E. California Blvd., Pasadena, CA 91125}
\email{carl@ipac.caltech.edu}

\author{Jeffrey L. Carlin \& Steven R. Majewski}
\affil{Department of Astronomy, University of Virginia, P.O. Box
400325, Charlottesville, VA 22904-4325}
%\email{jc4qn@mail.astro.virginia.edu}

\begin{abstract}

We have obtained radial velocity measurements for stars in two,
widely-separated fields in the Anticenter Stream. Combined with
SDSS/USNO-B proper motions, the new measurements allow us to establish
that the stream is on a nearly circular, somewhat inclined, prograde
orbit around the Galaxy.  While the orbital eccentricity is similar to
that previously determined for the Monoceros stream, the sizes,
inclinations, and positions of the orbits for the two systems differ
significantly. Integrating our best fitting Anticenter Stream orbit
forward, we find that it is closely aligned along and lies almost on
top of a stream-like feature previously designated the ``Eastern
Banded Structure''. The position of this feature coincides with the
apogalacticon of the orbit. We tentatively conclude that this feature
is the next wrap of the Anticenter Stream.

\end{abstract}



\section{Introduction}

With numerous stellar streams now known to inhabit the Galactic halo
\citep{ibata2001, maje2003, yann03, rocha2004, belokurov2006,
grillo2006a, grillo2006b, oden2001, grill2006b,
  grill2006c, grill2008} it appears that we are entering an
interesting new era in which we can hope gradually to unravel the
accretion events that are thought to have built up our Galaxy. As the
formation and dynamical evolution of tidal streams are strongly
affected by the Galactic potential, careful study of their
positions and motions will help us to refine our knowledge of both the
global and small-scale distribution of dark matter
\citep{murali99, johnston2002, johnston2005}. 

The Anticenter Stream (ACS) is one of the most visible features in
stellar maps (e.g. Belokurov et al. 2007, Grillmair 2006b - hereafter
G06) derived from the Sloan Digital Sky Survey (SDSS). The ACS extends
for some 65\arcdeg~ along the western edge of the survey area.
G06 used a matched-filter technique to
examine the detailed structure of the stream. Among the findings of
that work was that the stream appears to be made up of at least three
separate, relatively cold streams, possibly the remnants of
dynamically distinct components within the progenitor. Of particular
interest was the finding that, since the ACS apparently does not
itself pass through Monoceros, and since it lies some 15\arcdeg~ to
the west of the best estimate of Monoceros' orbit
\citep{penarrubia2005}, the two streams must be distinct. Using the
position of the stream on the sky and an estimated distance, G06 was
able to put limited constraints on the orbit of the stream. The chief
uncertainty in this estimate was the lack of radial velocity and
proper motion data.

In this paper we make progress on alleviating this uncertainty by
presenting new radial velocity measurements of stars in the ACS
(Section \ref{sec:observations}). We estimate stream velocities
and proper motions in Section \ref{sec:analysis} and put new constraints
on the orbit of the ACS in Section \ref{sec:orbit}.

\section{Observations}
\label{sec:observations}

Radial velocity measurements for stars selected from the SDSS were
carried out using the HYDRA multi-object spectrograph on the WIYN
telescope in February
2007.  Target stars were selected to have $18.6 < g < 20.1$ and $0.21 < g - r
< 0.43$, which places them on or near the color-magnitude sequence obtained
for the ACS by G06. The sample in one field was
further culled with proper motions from the \citet{cd2006} study of
Selected Area (SA) 76, which lies along the ACS; stars were selected to
have solar reflex-corrected proper motions consistent (to
within $\pm 2\sigma$) with motion along the north-south orientation of
the ACS. A total of 123 stars were observed in two $50\arcmin \times
50\arcmin$ fields: ACS-B centered at $(\alpha,\delta)_{2000} =
(124, 37.5)^{\circ}$, and ACS-C (SA
76) centered at $(\alpha,\delta)_{2000} = (125, 14.7)^{\circ}$. 

We used the 600@10.1 grating in first order, centered at 5400\AA, to
give a working wavelength range of 4000-6800\AA~at a dispersion of
1.397 \AA$~$pix$^{-1}$ and a spectral resolution of 3.35 \AA.  This
region was selected to include the H$\beta$, Mg triplet, Na D, and
H$\alpha$ spectral features, among others.  Integration times totaled
9x2400 seconds (6 hours) for each ACS field, though the ACS-C
observations were beset by high winds and poor ($\sim2.5\arcsec$)
seeing. Fifteen measurements of sky background were taken at random
positions in each field.

The HYDRA data were reduced using the procedures described in
\citet{cd2008}. Computed velocity errors range from 6 to 20 km
s$^{-1}$, with a mean of 11.5 km s$^{-1}$ for ACS-B. For ACS-C,
computed errors range from 4 to 21 km s$^{-1}$, with a mean of 9.3 km
s$^{-1}$. A more detailed description of the observations will be
presented in a forthcoming paper \citep{carlin2009}.

\section{Analysis}
\label{sec:analysis}

\begin{figure}
\epsscale{0.90}
\plotone{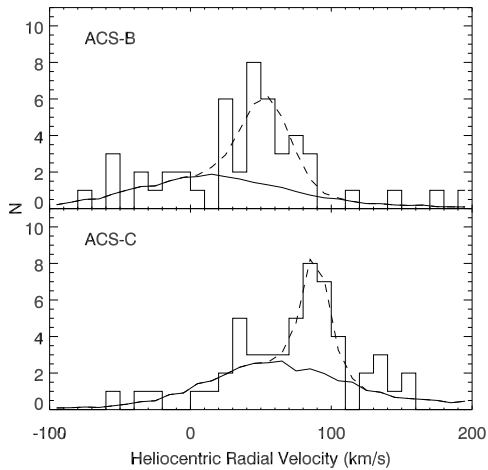}
\caption{Histograms of radial velocities measured in our
two fields. The smooth curves show the scaled distributions of
foreground stars as predicted by the Besancon model of the Galaxy. The
dashed curves show the Gaussian distributions that best fit the
stream peaks. The histograms are shown using 10 km s$^{-1}$-wide
velocity bins, though our analysis makes use of a variety of bin
widths.} 
\label{fig:histograms}
\end{figure}

Figure \ref{fig:histograms} shows velocity histograms for stars in the
two ACS fields. Also shown are the scaled radial velocity distributions
(see below) predicted by the Besancon model of the Galaxy
\citep{robin2003} in these directions over the magnitude and color
ranges of our targets.  The foreground distributions comprise over
10000 realizations and for practical purposes are free of random
errors.  For the restricted color and magnitude ranges applied to our
targets, the predicted line-of-sight velocity dispersions of field
stars are $\sim 70$ km s$^{-1}$.

Narrow velocity peaks are visible in both the ACS-B and ACS-C fields.
We test the similarity of the observed and model distributions of
radial velocities using a two-tailed Kolmogorov-Smirnoff test. For
ACS-B there is a $\sim 50$ km s$^{-1}$ difference in the peak
velocities of the two distributions, and we find that we can reject
the hypothesis that the observed and predicted distributions were
drawn from the same parent population at the 98.4\% signficance level.
For ACS-C the significance level of the null hypothesis is only 84\%
and we cannot rule out that the distributions were drawn from the same
parent population.  This is due to the higher mean velocity of
foreground stars in this direction ($\sim 60$ km s$^{-1}$), and to the
presence of the two apparent peaks (at 35 and 90 km s$^{-1}$) that
bracket that mean velocity.  The surface density of SDSS stars that
match our color and magnitude selection criteria is 25\% lower in the
ACS-C field than in ACS-B, but the strength of the stream (as measured
from the filtered star counts of G06) is also about 35\% lower in
ACS-C. As a percentage, sample contamination should therefore be
similar in the two fields. Given that we observed the same total
number of stars in each field, and that we observed only a fraction of
the total number of stars available, we expect similar numbers of
stream stars to have been selected in each field. The more
pronounced ACS-C peak at $\sim 90$ km s$^{-1}$ is clearly at odds with
the predicted foreground distribution and the integrated number of
stars in the peak (see below) approximately matches the number of
stream stars in the ACS-B peak. We consequently adopt the 90 km
s$^{-1}$ peak as being due to ACS stars.

\begin{deluxetable}{lccccr}
\scriptsize
\tablecaption{Anticenter Stream Velocities and Proper Motions}
\tablecolumns{6}
\tablewidth{0pc}
\tablehead{
\multicolumn{1}{c} {Field} &
\multicolumn{1}{c} {V$_s$} &
\multicolumn{1}{c} {$\sigma_s$} &
\multicolumn{1}{c} {N$_s$} &
\multicolumn{1}{c} {$\mu_{\alpha}\cos{\delta}$} &
\multicolumn{1}{c} {$\mu_{\delta}$} \\
\multicolumn{1}{c} {} &
\multicolumn{1}{c} {km s$^{-1}$} &
\multicolumn{1}{c} {km s$^{-1}$} &
\multicolumn{1}{c} {} &
\multicolumn{1}{c} {mas yr$^{-1}$} &
\multicolumn{1}{c} {mas yr$^{-1}$} }
\startdata
ACS-B & $53.1 \pm 7.5$ & $16.1 \pm 5.9$ & 22 & $0.29 \pm 0.95$ & $-1.94 \pm 0.95$ \\
ACS-C & $88.8 \pm 5.0$ & $6.9 \pm 3.6$ & 16 & $0.67 \pm 0.81$ & $0.73 \pm 0.80$ 
\enddata
\end{deluxetable}

For each field the observations are modeled as a sum of the scaled,
Besancon distribution of foreground stars and a single Gaussian with mean
velocity $V_s$, velocity dispersion $\sigma_s$, and number of stream
stars $N_s$.  The four fitting parameters are solved for in a least
squares sense, and the results are given in Table 1.  We use bin sizes
of 3, 5, 10, and 15 km s$^{-1}$, beyond which the stream peaks become
too undersampled to give reliable results.  The fitted parameters are
very nearly identical in all cases (with a dispersion of $\sim 0.8$ km
s$^{-1}$), and we adopt a binsize of 5 km s$^{-1}$ for the results
given in Table 1. The uncertainties are estimated by constructing 1000
realizations of the data, generating Gaussian deviates for each
velocity measurement using the individual measurement errors, and
rebinning the results. The velocity bin walls are also randomly offset
by from 1 to 4 km s$^{-1}$, respectively.  The uncertainties are 
estimated by measuring the standard deviations in the resulting best-fit
parameters.

The observed stream star velocity dispersions are the
convolution of the stellar velocity dispersions and the uncertainties
in the measurements. Correcting the measured velocity dispersions by
quadrature subtraction of the mean uncertainties in each field, we
find stellar velocity dispersions of 14.9 and 5.9 km$^{-1}$ for ACS-B
and ACS-C, respectively. The latter is consistent with a very
cold stream and suggests that we are primarily sampling one of the
nearly parallel sub-streams or ``tributaries'' found by G06. The
dispersion in the ACS-B field may indicate that we are measuring stars
in more than one tributary or more than one orbital wrap
(see below).

All of our target stars also have proper motion measurements based on
a SDSS--USNO-B positional comparison \citep{monet2003, munn2004}.
Though the individual measurement uncertainties
(typically 4 mas yr$^{-1}$) are quite sizable, averaging over a
selected sample can be used to put limits on the motions of the
stream. While the original SDSS proper motion catalog had small errors
in the R.A.  component \citep{ivezic2008}, we use corrected proper
motions kindly provided by J. Munn.  We estimate the mean stream
proper motion in each field by averaging over only those stars with
radial velocities within $1\sigma$ of the best-fit stream velocities.
The weighted average proper motions are given in Table 1.

\section{A New Orbit for the Anticenter Stream}
\label{sec:orbit}

G06 estimated the ACS to lie at a distance of $8.9 \pm 0.2$ kpc. We
constrain the path of the stream by selecting 30 fiducial points that
trace the tributary that passes through ACS-B and ACS-C.  There is
some uncertainty in this procedure because the different dynamical
components of the stream are not uniformly strong along their lengths,
and it is possible to confuse the different tributaries as they blend,
cross one another, or fade entirely. Based on the scatter in the
estimated fiducial points about a second order polynomial fit, we
assign a positional uncertainty of 0.35\arcdeg~ to each point.

\begin{figure}
\epsscale{.8}
\plotone{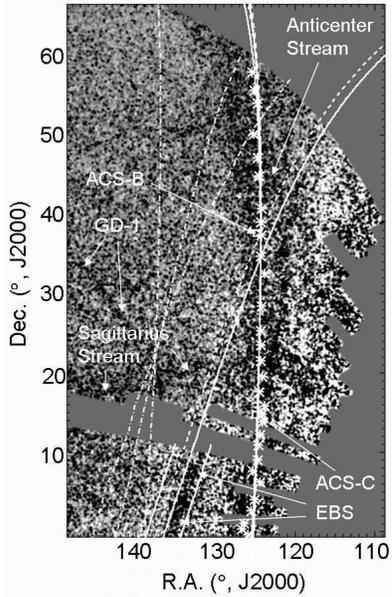}
\caption{Our best-fitting orbit projections overlaid on the ACS. The
underlying image is a surface density map of the western
portion of the SDSS survey area, filtered for an M 13-like stellar
population at a distance of 9 kpc \citep{grill2006c}. A low-order
polynomial surface fit has been subtracted to remove large-scale
non-uniformities, and the result has been smoothed with a Gaussian
kernel with $\sigma = 0.2\arcdeg$. The solid curves show the first and
second wraps for Model A (unconstrained by proper motions) along with
their 1$\sigma$ limits in the vicinity of the EBS, while the dashed
lines show orbit projections for Model B (constrained using the proper
motions in Table 1). The dash-dot curves show the nearby portions of
the best-fitting prograde orbits for the Monoceros stream as
determined by \citet{penarrubia2005}. The asterisks show the fiducial
points used to constrain the position of the ACS, and the positions of
ACS-B and ACS-C are indicated.}
\label{fig:sky}
\end{figure}

Using the Galactic model of \citet{allen91} (which assumes a spherical
halo potential), we generate test particle orbits and then use
$\chi^2$-minimization to simultaneously fit the radial velocities,
proper motions, distance, and the apparent path of the stream.  We use a
downhill simplex method, integrating orbits and comparing positions
and velocities at each step. As we use 30 sky position measurements
and only two velocity measurements, we increase the relative weight of
the velocity and proper motions measurements by a factor of 30/2 in
the computation of $\chi^2$.

\begin{figure}
\epsscale{1.0}
\plotone{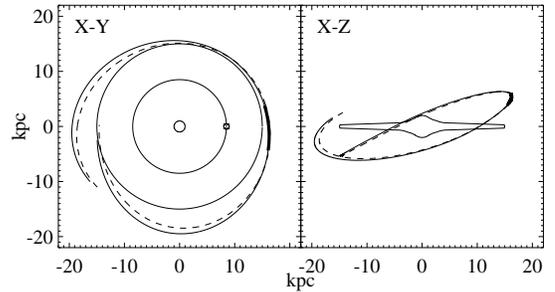}
\caption{Our best-fitting orbits in right-handed, Cartesian Galactic coordinates. The Sun's position is shown as the small open circles at R = 8.5 kpc. The solid line corresponds to Model A while the dashed line shows Model B. The thick portion of Model A shows the section of the ACS visible in the SDSS.} 
\label{fig:orbit}
\end{figure}

In Figure \ref{fig:sky} we show the projected paths of the
best-fitting prograde orbits, both with (Model B) and without (Model
A) proper motion contraints. The $1\sigma$ uncertainties in the
integrated orbits are computed by applying Gaussian deviates to the
radial velocities, proper motions, sky positions, and distance
measurements using their estimated individual uncertainties.  We
construct 100 such realizations and recompute the best-fitting orbit
in each case. The fit parameters and their estimated uncertainties are
given in Table 2. The orbits are shown in Cartesian Galactic coordinates in
Figure \ref{fig:orbit}.

For the case where proper motions are allowed to be free parameters,
the best-fit proper motions given in Table 2 lie within $1\sigma$ of the 
average values in Table 1 with the exception of $\mu_\delta$ for
ACS-B, which lies $3\sigma$ from its measured value. If the measured
proper motions are used to constrain the fit, this departure is
reduced to $1.8\sigma$. At this point we cannot say whether the
difference in $\chi^2$ in Table 2 is due to possible errors
in the proper motion measurements or to the inappropriateness of our
relatively simple Galactic potential.

Unconstrained by proper motion measurements, a retrograde orbit
predicts $\mu_\delta = $ -10.0 mas yr$^{-1}$ and -10.1 mas yr$^{-1}$
for ACS-B and ACS-C, respectively. For stars with radial velocities
within $1\sigma$ of the stream velocities in Table 1, only two (ACS-B)
and zero (ACS-C) have $\mu_\delta < -10 $ mas yr$^{-1}$.  The mean
proper motions disagree with the retrograde orbit model at the 8 and
$14\sigma$ levels for the ACS-B and ACS-C fields, respectively. We
conclude with high confidence that the stars in the ACS
are orbiting the Galaxy in a prograde fashion.

Our computed orbital parameters predict that apogalacticon occurs at
($\alpha,\delta$) = (294\arcdeg, 57\arcdeg; Model A) or (293\arcdeg,
45\arcdeg; Model B), well outside the current limits of the SDSS. On
the other hand, integrating the orbits forward reveals that the
subsequent apogalactica occur at ($\alpha,\delta$) = (141\arcdeg,
-10.3\arcdeg, Model A) or (134\arcdeg, 5.1\arcdeg, Model B), which are
within or nearly within the field shown in Figure \ref{fig:sky}. This
second wrap of the orbit is closely aligned with, and lies almost on
top of, the stream-like feature identified by G06 as the ``Eastern
Banded Structure'' (EBS). The apparent agreement between the predicted
orbits and the position of the EBS is highly suggestive.  The point
where the two wraps cross (in projection) is predicted to occur at an
R.A. of 125\arcdeg and declination of between 34\arcdeg~ and
37\arcdeg, very close to the ACS-B field. Could we have inadvertantly
sampled stars from both the first and second wraps of the stream and
could this account for the higher velocity dispersion seen in the
ACS-B field?  The distance of the second wrap at this point is
predicted to be $\sim 15$ kpc, or about 1.1 mag fainter than the ACS
itself. The turn-off and sub-giant branches of the two wraps should
therefore overlap, and it is certainly possible that some of our
fainter targets could belong to the second wrap.  The predicted radial
velocity of the second wrap at this point is -15 to -17 km s$^{-1}$.
While examination of Figure \ref{fig:histograms} does not reveal an
obvious peak in the velocity distribution at this point, only a small
number of second-wrap stars blended in with primary wrap stars would
be sufficient to broaden the velocity peak.  On the other hand, if we
divide our observations into two, roughly equal samples of stars with
$g < 19.6$ and $g > 19.6$ respectively, we find that the best-fit
velocity dispersion (after accounting for differing mean velocity
errors) is larger for the brighter stars (17 km s$^{-1}$) than it is
for the fainter stars (11 km s$^{-1}$).  We conclude that 
that second wrap stars are unlikely to account for the
larger velocity disperion in ACS-B, and that the observed
velocity dispersion is either intrinsic or due to the overlap of
multiple tributaries in the primary wrap.

The strongest concentration of stars in the EBS occurs at
($\alpha,\delta$) = (133.8\arcdeg, 3.4\arcdeg) and would be a
favorable field in which to measure radial velocities for this stream.
At this declination, both our models for the second wrap of the ACS
predict a heliocentric radial velocity of 20 km s$^{-1}$, and proper
motions ($\mu_{\alpha} \cos{\delta}, \mu_{\delta}$) = (-0.8, -0.14) mas
yr$^{-1}$.  If these predictions are borne out by future
observations, the case for a physical association between the ACS and
the EBS will be considerably strengthened.

These orbit determinations do not agree with the preliminary estimate
of G06, underscoring the fact that projected position and distance
alone are insufficient to provide reliable constraints even for
relatively long streams.  A comparison of the orbital parameters with
those of \citet{penarrubia2005} for their best-fitting prograde orbits
for the Monoceros stream reveals some interesting similarities. While
these authors put the apogalacticon of Monoceros some 3 kpc farther
out and the orbital inclination some 5\arcdeg~ larger than what we
find for the ACS, the orbital eccentricity they find
(0.1) is essentially identical to that found here. In Figure \ref{fig:sky} we show
the projected paths of the nearby portions of their best-fit prograde
orbits for $q_h = 0.6, 0.7,$ and 0.8. The western-most projection
($q_h = 0.8$) bears some similarity to the second wrap of the
ACS, though the sky position remains offset some
8\arcdeg~ to the east. Given this near agreement, it is interesting to
consider whether backward integration of the Monoceros orbit another
full wrap may provide a reasonable match to the position of the
ACS itself.

\section{Conclusions}
\label{sec:conclusions}

Radial velocity and proper motion measurements in two fields of the
ACS show that the stream is in a prograde orbit. 
Integrating the orbit forward we find that the next
wrap of the ACS is aligned along and lies almost on top
of the stream known as the EBS.

%JLC: When you mention ``more accurate proper motions'', should we 
% refer to Casetti-Dinescu et al. 2006 and/or Carlin et al. 2009
% better PMs for SA76?
Further refinement of the orbit will require additional velocity
measurements, more accurate proper motions, and a more realistic model
of the Galactic potential. In this respect, planned spectroscopic
surveys such as SEGUE2, APOGEE, and LAMOST, and proper motions from
GAIA, LSST, and the SIM-Lite will significantly advance our
understanding of this particular stream's motion and origin and,
ultimately, the shape of the Galactic potential.

\acknowledgments

We are indepted to J. Munn for providing corrected SDSS-USNO-B proper
motions in advance of the release of SDSS DR7. JLC and SRM acknowlege
NSF grants AST-0307851 and AST-0807945.

{\it Facilities:} \facility{NOAO}.

\begin{deluxetable}{lccccccccccc}
\tabletypesize{\scriptsize}
%\rotate
\tablecaption{Best-Fit Orbit Parameters}
\tablecolumns{12}
\tablewidth{0pc}
\tablehead{
\multicolumn{1}{c} {Model} &
\multicolumn{2}{c} {$V_R$} &
\multicolumn{4}{c} {Proper Motions} &
\multicolumn{1}{c} {$R_{peri}$} &
\multicolumn{1}{c} {$R_{apo}$} &
\multicolumn{1}{c} {$e$} &
\multicolumn{1}{c} {$i$} &
\multicolumn{1}{c} {$\chi^2$} \\
\multicolumn{1}{c} {} & 
\multicolumn{1}{c} {ACS-B} &
\multicolumn{1}{c} {ACS-C} &
\multicolumn{2}{c} {ACS-B} &
\multicolumn{2}{c} {ACS-C} &
\multicolumn{5}{c} {} \\
\multicolumn{3}{c} {} &
\multicolumn{1}{c} {$\mu_\alpha$ cos($\delta$)} &
\multicolumn{1}{c} {$\mu_\delta$} &
\multicolumn{1}{c} {$\mu_\alpha$ cos($\delta$)} &
\multicolumn{1}{c} {$\mu_\delta$} &
\multicolumn{5}{c} {} \\
\multicolumn{1}{c} {} &
\multicolumn{2}{c} {km s$^{-1}$} &
\multicolumn{2}{c} {mas yr$^{-1}$} &
\multicolumn{2}{c} {mas yr$^{-1}$} &
\multicolumn{1}{c} {kpc} &
\multicolumn{1}{c} {kpc} &
\multicolumn{1}{c} {} &
\multicolumn{1}{c} {\arcdeg} &
\multicolumn{1}{c} {}  }
\startdata
A & $51.5 \pm 7.0$ & $89.1 \pm 5.5$& $0.80 \pm 0.04$& $-0.11 \pm 0.34
$ & $ 0.64 \pm 0.03$ & $0.67 \pm 0.35 $ & $15.5 \pm 0.9$ & $19.8 \pm
2.2$ & $0.12 \pm 0.07 $& $20.1 \pm 0.4 $ & 1.2 \\
B & $ 47.8 \pm 7.1 $ & $90.3 \pm 5.2 $&$0.78 \pm 0.16 $  & $-0.27 \pm 0.30$
& $0.61 \pm 0.18$ & $ 0.40 \pm 0.32 $ & $15.4 \pm 1.1$ & $19.0 \pm
1.9$ & $0.10 \pm 0.07 $& $20.1 \pm 0.7 $ & 2.3 
\enddata
\end{deluxetable}

\clearpage

%% the file, uncomment the \plotone, \plottwo, and
%% If you need a layout that cannot be achieved with \plotone or
%% \plottwo, you can invoke the graphicx package directly with the
%% \includegraphics command or use \plotfiddle. For more information,
%% please see the tutorial on "Using Electronic Art with AASTeX" in the
%% documentation section at the AASTeX Web site,
%% http://www.journals.uchicago.edu/AAS/AASTeX.
%%
%% The examples below also include sample markup for submission of
%% supplemental electronic materials. As always, be sure to check
%% the instructions to authors for the journal you are submitting to
%% for specific submissions guidelines as they vary from

%% This example uses \plotone to include an EPS file scaled to
%% 80% of its natural size with \epsscale. Its caption
%% has been written to indicate that additional figure parts will be
%% available in the electronic journal.


\begin{thebibliography}{}

\bibitem[Allen \& Santillan (1991)]{allen91} Allen, C., \& Santillan,
A. 1991, {\it Rev. Mex. Astron. Astrofis.}, 22, 255

\bibitem[Belokurov et al. (2006)]{belokurov2006} Belokurov, V., et
al. 2006b, \apjl, 642, 137

\bibitem[Belokurov et al. (2007)]{belokurov2007} Belokurov, V., et
al. 2007, \apj, 658, 337

\bibitem[Carlin et al. (2009)]{carlin2009} Carlin, J. L, Grillmair,
C., \& Majewski, S. R. 2009, in prep.

\bibitem[Casetti-Dinescu et al.(2008)]{cd2008} 
Casetti-Dinescu, D.~I., Carlin, J.~L., Girard, T.~M., Majewski, S.~R., 
Pe{\~n}arrubia, J., \& Patterson, R.~J.\ 2008, \aj, 135, 2013 

\bibitem[Casetti-Dinescu et al.(2006)]{cd2006} 
Casetti-Dinescu, D.~I., Majewski, S.~R., Girard, T.~M., Carlin, J.~L., van 
Altena, W.~F., Patterson, R.~J., \& Law, D.~R.\ 2006, \aj, 132, 2082 

\bibitem[Grillmair \& Dionatos (2006a)]{grillo2006a} Grillmair,
C. J., \& Dionatos, O. 2006a, \apjl, 641, 37

\bibitem[Grillmair \& Dionatos (2006b)]{grillo2006b} Grillmair, C. J.,
\& Dionatos, O. 2006b, \apj, 643, L17

\bibitem[Grillmair (2006a)] {grill2006b} Grillmair, C. J. 2006a, \apjl, 645, L37

\bibitem[Grillmair (2006b)]{grill2006c} Grillmair, C. J. 2006b, \apjl,
651, L29

\bibitem[Grillmair (2008)]{grill2008} Grillmair, C. J. 2008, \apj,
submitted.

\bibitem[Ibata et al. (2001)]{ibata2001} Ibata, R., Lewis, G.~F.,
Irwin, M., Totten, E., \& Quinn, T. 2001, \apj, 551, 294

\bibitem[Ivezic et al. (2008)]{ivezic2008} Ivezic, Z. et al. 2008,
\apj, 684, 287

\bibitem[Johnston et al. (2002)]{johnston2002} Johnston, K.~V.,
Spergel, D.~N., \& Hadyn, C. 2002, \apj, 570, 656

\bibitem[Johnston et al. (2005)]{johnston2005} Johnston, K.~V., Law,
D.~R., \& Majewski, S.~R. 2005, \apj, 619, 800

\bibitem[Majewski et al. (2003)]{maje2003} Majewski, S. R., Skrutskie,
M. F., Weinberg, M. D., \& Ostheimer, J. C. 2003, \apj, 599, 1082

\bibitem[Monet et al. (2003)]{monet2003} Monet, D. G. et al. 2003,
\aj, 125, 984 

\bibitem[Munn et al. (2004)]{munn2004} Munn, J. A. et al. 2004, \aj,
127, 3034

\bibitem[Murali \& Dubinski (1999)]{murali99} Murali, C., \& Dubinski,
J. 1999, \aj, 118, 911

\bibitem[Odenkirchen et al. (2001)]{oden2001} Odenkirchen, M., et
al. 2001, \apjl, 548, 1650

\bibitem[Penarrubia et al. (2005)]{penarrubia2005} Penarrubia, J,
Martinez-Delgado, D., Rix, H. W., Gomez-Flechoso, M. A., Munn, J.,
Newberg, H., Bell, E. F., Yanny, B., Zucker, D., \& Grebel,
E. K. 2005, \apj, 626, 128

\bibitem[Robin et al. (2003)]{robin2003} Robin, A. C., Reyle, C.,
Derriere, S., \& Picaud, S. 2003, \aa, 409, 523

\bibitem[Rocha-Pinto et al. (2004)]{rocha2004} Rocha-Pinto, H.~J.,
Majewski, S.~R., Skrutskie, M.~F., Crane, J.~D., \& Patterson,
R.~J. 2004, \apj,  615, 732

\bibitem[Yanny et al. (2003)]{yann03} Yanny, B. et al. 2003, \apj, 588, 824

\end{thebibliography}
\end{document}